\documentclass{article}
\usepackage{spconf,amsmath,graphicx}

\usepackage{enumitem}
\setlist{nosep, leftmargin=14pt}

\usepackage{mwe} 

\usepackage{comment}

\title{Self-Learned Kernel Low Rank Approach to Accelerated High Resolution 3D Diffusion MRI}
\name{Abhijit Baul$^{1}$, Nian Wang$^{2}$, Choyi  Zhang$^{3}$, Leslie Ying$^{3,4}$, Yuchou Chang$^{5}$, Ukash Nakarmi$^{6}$}
\address{$^{1}$Department of Electrical Engineering and Computer Science, University of Texas Rio Grande Valley \\ 
$^{2}$Department of Radiology and Imaging Sciences, Indiana University School of Medicine\\
$^{3}$Department of Electrical and Computer Engineering, University at Buffalo\\
$^{4}$Department of Biomedical Engineering, University at Buffalo\\
$^{5}$Department of Computer and Information Science, University of Massachusetts-Dartmouth\\
$^{6}$ Department of Computer Science and Computer Engineering, University of Arkansas}

\begin{document}
%
\maketitle
\begin{abstract}
Diffusion Magnetic Resonance Imaging (dMRI) is a promising method to analyze the subtle changes in the tissue structure. However, the lengthy acquisition time is a major limitation in the clinical application of dMRI. Different image acquisition techniques such as parallel imaging, compressed sensing,  has shortened the prolonged acquisition time but creating high-resolution 3D dMRI slices still requires a significant amount of time. 
In this study, we have shown that high resolution 3D dMRI can be reconstructed from the highly undersampled k-space and q-space data using a Kernel Low Rank method. Our proposed method has outperformed the conventional CS methods in terms of both image quality and diffusion maps constructed from the diffusion-weighted images.   

\end{abstract}
\section{Introduction}
\label{sec:intro}

Diffusion Magnetic Resonance Imaging (dMRI) has immense clinical application in tissue analysis, especially in grey and white matter study. The method is continuously evolving, beginning from Diffusion Tensor Imaging (DTI)~\cite{DTI} to recent High Angular Resolution Diffusion Imaging (HARDI) and Diffusion Spectrum Imaging (DSI)~\cite{DSI} which provides the complex fiber orientation in the tissue. Several important maps like Mean Diffusivity (MD) and Fractional Anisotropy (FA) can be generated from 3D dMRI, providing essential details about the micro structure of the tissue. One major limitation is the requirement of several diffusion-weighted images (DWIs) in different diffusion directions corresponding to so called q-space; for example, the DTI requires at least six DWIs to estimate the 3D trajectory of the diffused water molecules, while DSI and HARDI require more than thirty DWIs along with different directions. The improvement in the data acquisition process like parallel imaging~\cite{parallel} and CS-based reconstruction~\cite{sparsemri} has reduced the overall acquisition time. In the case of dMRI, some acceleration  methods perform the undersampling in the diffusion encoding directions and, some methods~\cite{wang} perform undersampling in the k-space only, ignoring the high correlations among the DWIs. The Kernel Low Rank (KLR) method was introduced by~\cite{kpca1} and~\cite{kpca_base} applied that to reconstruct high resolution 3D dMRI images from undersampling in both k-space and q-space.  However, the training process described by~\cite{kpca_base} requires low resolution dWIs, obtained from the central region of the k-space to enforce Kernel Low Rankness in the reconstruction process. 

In this study, we demonstrate that kernel principal components can be directly estimated from the highly undersampled k-space and q-space data. Previous studies on dMRI mainly focused on the brain region; here, we performed our experiments on the meniscus region of the knee joint. Our main contributions are : 1) Developed a Kernel Low Rank method that do not require low-resolution images, 2) Acquired High Resolution ground truth data with several diffusion directions and investigated the effect of i) undersampling in k-space only and ii) undersampling in both k-space and diffusion direction for meniscus dMRI, 3) Studied the efficiency of Kernel-low-rank technique in meniscus diffusion MRI.
\section{Background}
In case of fully sampled dMRI, the image data $D$ consists of $K_x\times K_y\times K_z \times K_q $ dimension image tensor where $K_x\times K_y\times K_z$ is the volume dimension and $K_q$ is the numbers of diffusion directions. When the data is under sampled during data acquisition, undersampling could be done in the k-space as well as in q-space (having fewer diffusion directions). However, reconstruction of image data from such undersampled data is ill-posed and results in aliased images. 
Similar to low rank and sparse methods ~\cite{sparsemri,linearpca} that reconstruct images from umdersampled k-space data, the Kernel Low Rank (KLR) also maps the data into a low dimensional embedding but unlike in low rank and sparse techniques, KLR applies a non-linear transformation to define the new representation space and enforces low rankness in the kernel space. Suppose, $\Phi$ be the non-linear mapping function and $x$ be the input, then the eigen vector $\textbf{v}$ of the covariance matrix $\Omega$ formed by $\overline{\Phi}(x)$ spans the kernel feature space and can be described as,
\begin{equation}
\label{eq: eigv}
    \begin{aligned}
     \textbf{v}=\sum_{i=1}^{N} \alpha_{i}  \overline{\Phi}(x)
     \end{aligned}
\end{equation}
where $\overline{\Phi}(x)$ is the mean centered data and $N$ is the number of  training signals. The coefficients $\alpha_{i}$ can be obtained by solving the eigen equations, $\lambda\textbf{v}$ = $\Omega\textbf{v}$ or equivalently,
\begin{equation}
\label{eq: k}
    \begin{aligned}
     K^c\alpha=\lambda\alpha
     \end{aligned}
\end{equation}
where $K^c$ is the centered kernel matrix and $k(x_i,x_j) = <\Phi(x_i),\Phi(x_j)>$ is the so called kernel function. The Kernel Low Rank representation of training and test signals $x_t$ can be achieved by projecting the data vectors $x_t$ to the first few principal kernel Eigen Vectors $v$. However, such representation requires data vectors to be projected back to the image space after low-rank enforcement which gives rise to the pre-imaging problem~\cite{kpca1}. A direct solution to the pre-imaging problem can be acquired by choosing an invertible kernel functions\cite{kpca1}.

\section{Methodology}
\subsection{Data Collection and Compliance with Ethical Standards}
Porcine menisci were harvested from the knee joints of skeletally mature pigs obtained from a local abattoir. All experiments in data acquisition procedures were approved by the Animal Use and Care Committee at the university. The dMRI Field of View (FOV) = 48*34.13*34.13, resolution = 360x256x256. There were eight non-diffusion weighted images ($b0$) and eighty-one diffusion direction and b = 1000$s/mm^2$.

\subsection{Proposed Method}
The reconstruction of dMRI from the undersampled data can be divided into three parts~\cite{kpca1}, i) Estimation of the diffusion bases (Kernel Eigen Vectors) from the low resolution training data, ii) Projection of the undersampled testing data along the diffusion direction and, iii) Reconstruction of the projected data into input space by applying pre-imaging and iterative data consistency in the k-space. In~\cite{kpca1}, the low resolution training images were obtained from the central region of the k-space. Our approach uses the same three steps as in~\cite{kpca1} except that instead of using training data vectors from low-resolution images, we hypothesize that data vectors from the undersampled images itself can be used as training data. We argue that our hypothesis is valid because essentially, we are using only the first few Kernel Eigen Vectors for reconstruction and the characteristics of these principal Kernel Eigen Vectors are unaffected by the artifacts introduced because of aliasing in the under sampled images. This is mainly because the artifacts mostly contribute to the characteristics (estimation) of minor Kernel Egen vectors not used in the reconstruction. We have verified this argument for a particular polynomial kernel, $k(x_i,x_j) = (<x_i,x_j>+b)^c$, and have shown that reconstruction can be done directly from the undersampled data without the use of low-resolution images. 

\section{Experiments}

\subsection{Training and Reconstruction}
In the beginning, we estimate the Kernel Eigen Vectors (diffusion bases)  that spans the diffusion space. To do that, we chose 2,000 random voxels $(k_x,k_y)$ locations along $k_q$ direction  from undersampled aliased images instead of low-resolution images such that our data vectors $x_t$ in \ref{eq: eigv} is of length $K_q \times 1$. Since we learn these diffusion bases from undersampled images itself we term this as self-learned KLR.  After that, we projected each voxel of undersampled data along the diffusion bases. Finally, the reconstruction using the projected data was done in two steps, similarly as in ~\cite{kpca1}, i) pre-imaging the projected data back to input space, known as and, ii) iterative data consistency to ensure that the reconstructed dMRI remains close to the input data~\cite{kpca_base}. We repeated such process for each $k_z$ to finally achieve a high-resolution 3D dMRI. The undersampling was done retrospectively in k-space at different reduction factors from 2-8 and in q-space by taking only 45 diffusion directions instead of 89. 


%
%
%


\subsection{Results and Discussion}
 We adopted the undersampling pattern described by~\cite{kpca_base}, along with phase encoding and slice encoding direction. We sampled the central k-space with a fixed radius, but the outer k-space was randomly sampled based on a probability function. We compared our reconstruction method with a conventional CS method~\cite{sparsemri} using wavelet and total variation sparsity priors along with spatial and diffusion directions respectively. We also compared the reconstruction using k-space and q-space undersampling i.e. using only 45 diffusion directions and k-space undersampling factors of 2-8. We observed that KLR method is much superior to conventional CS approaches when we have undersampled along the q-space.

\begin{figure}[ht]
 \centering
 \includegraphics[width=1.\columnwidth]{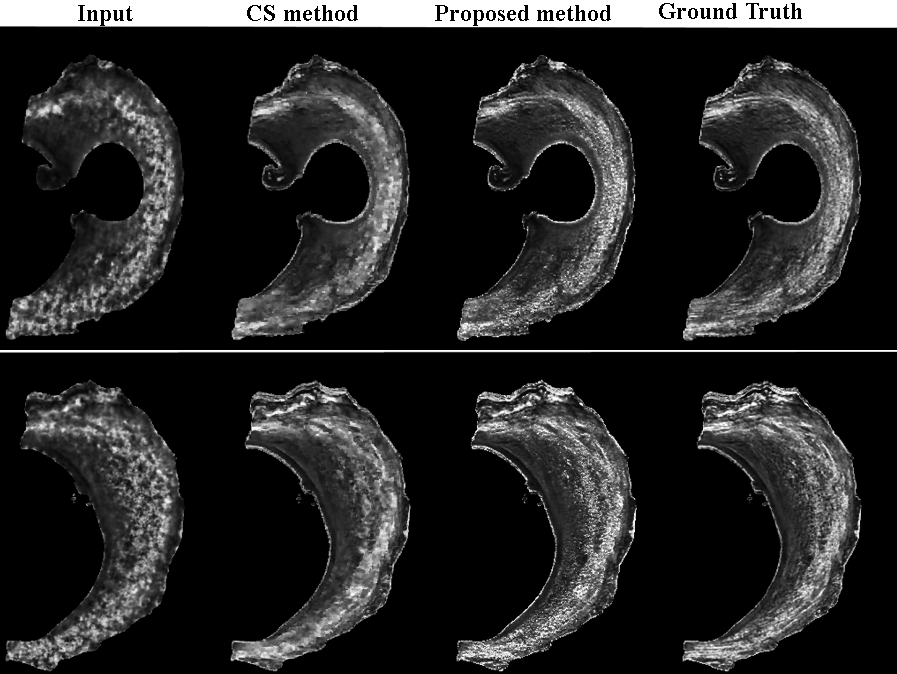}
 \caption{Example of the FA obtained from k-space undersampling (AF = 8).}
\label{fig:fig2}
\end{figure}
\begin{figure}[!ht]
 \centering
 \includegraphics[width=1.\columnwidth]{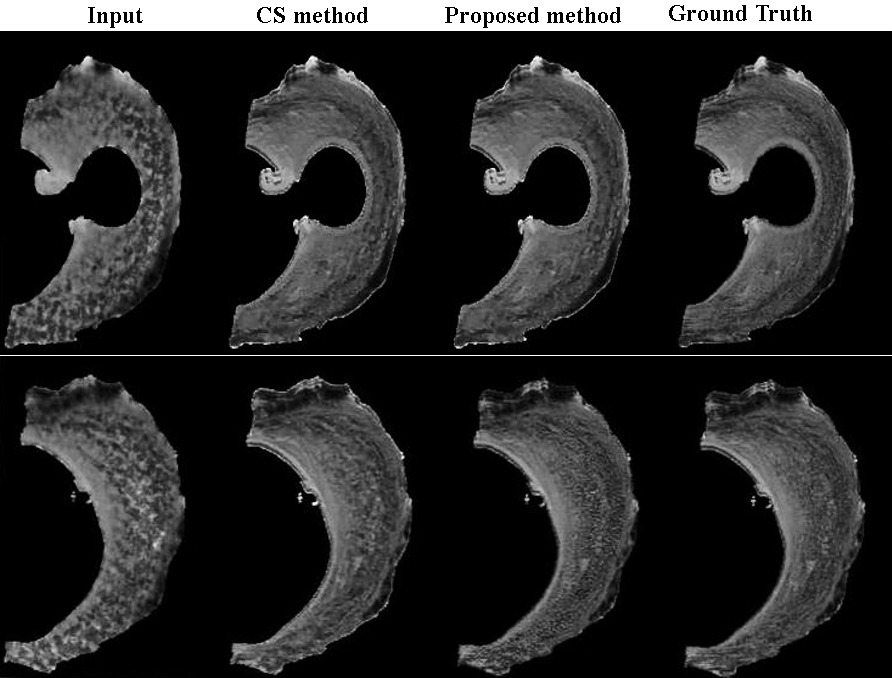}
 \caption{Example of the MD obtained from k-space undersampling (AF = 8).}
\label{fig:fig3}
\end{figure}
\begin{figure}[!ht]
 \centering
 \includegraphics[width=1.\columnwidth]{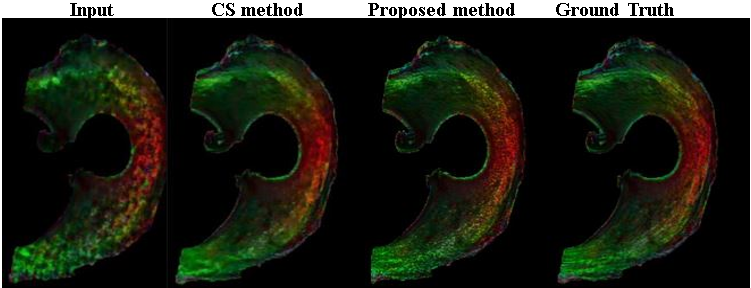}
 \caption{Example of the Colormap obtained from k-space undersampling (AF) = 8.}
\label{fig:fig4}
\end{figure}

\begin{figure}[!ht]
 \centering
 \includegraphics[width=1.\columnwidth]{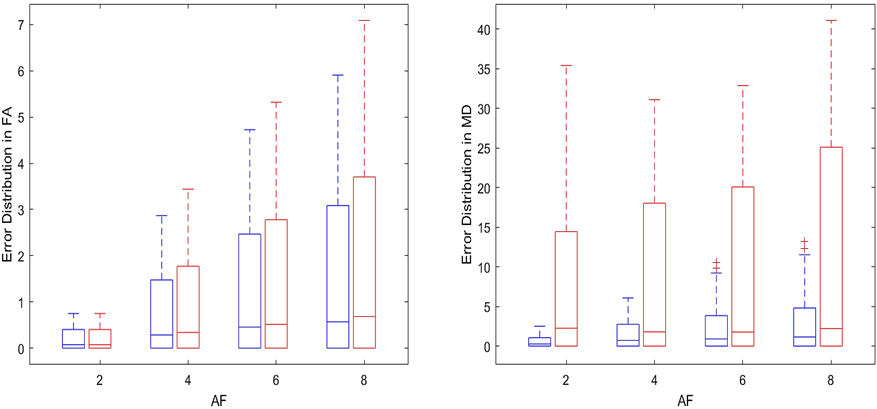}
 \caption{Error distribution in FA and MD (only k-space undersampling).}
\label{fig:plot 1}
\end{figure}

\begin{figure}[!ht]
 \centering
 \includegraphics[width=1.\columnwidth]{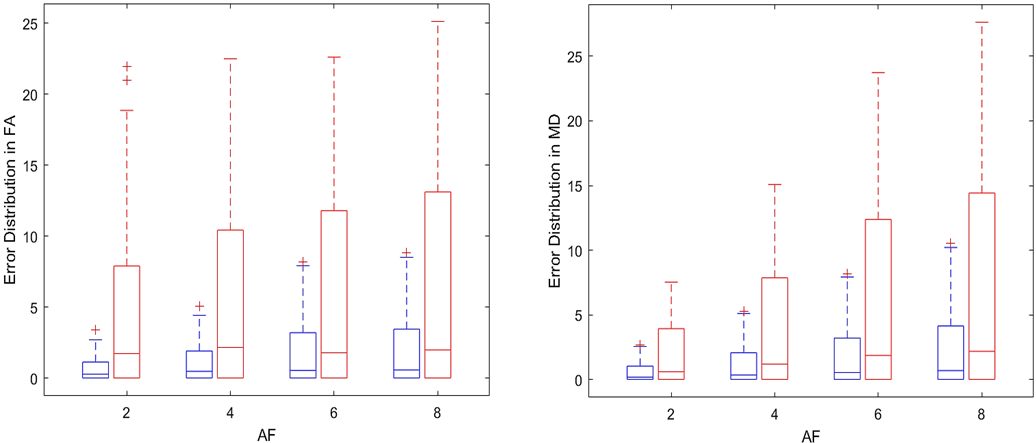}
 \caption{Error distribution in FA and MD (both k-space and q-space undersampling).}
\label{fig:plot 3}
\end{figure}

\begin{figure}[!ht]
 \centering
 \includegraphics[width=1.\columnwidth]{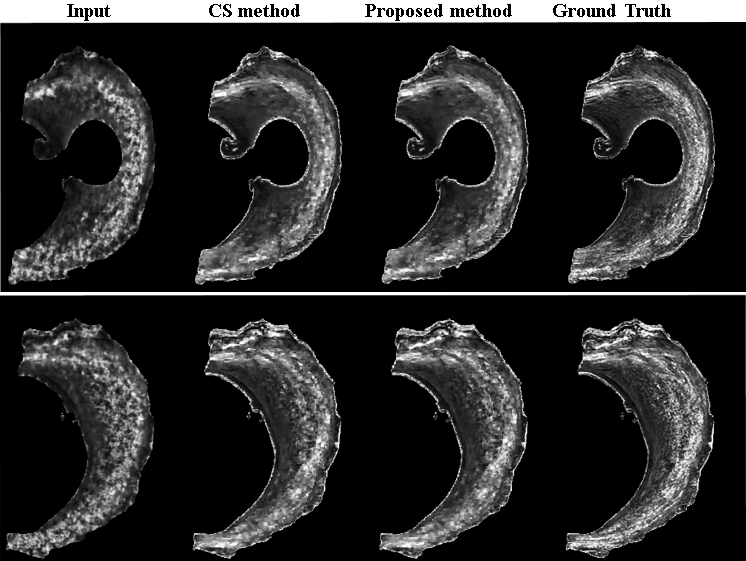}
 \caption{Example of the FA obtained from both k-space and q-space undersampling (AF = 8).}
\label{fig:fig5}
\end{figure}

\begin{figure}[!ht]
 \centering
 \includegraphics[width=1.\columnwidth]{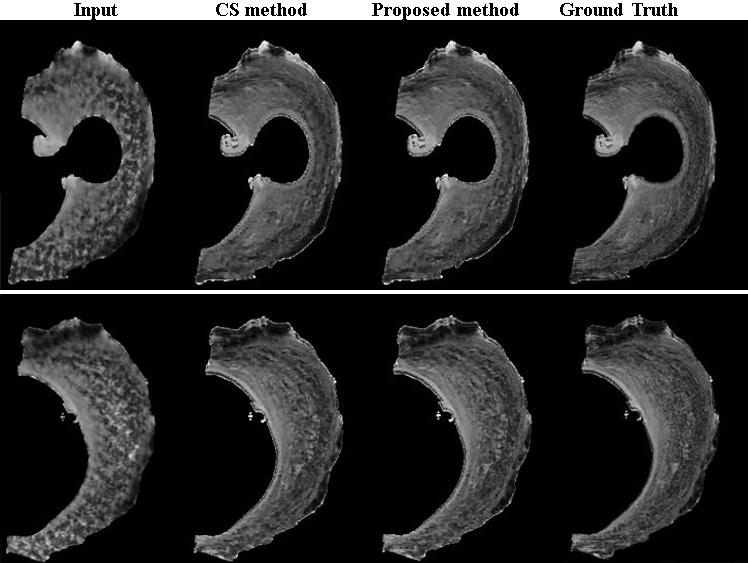}
 \caption{Example of the MD obtained from both k-space and q-space undersampling (AF = 8).}
\label{fig:fig6}
\end{figure}

To evaluate the quality of the reconstructed diffusion-weighted images, we used two widely used metrics normalized mean squared error (NMSE) and peak signal to noise ratio (PSNR). To evaluate the study of diffusion characteristics we used two important metrics used in Diffusion-weighted Imaging, i) Fractional Anisotropy (FA) and Mean Diffusivity (MD). FA provides information about the directionality of molecular displacement by diffusion while MD shows the average magnitude of molecular displacement by diffusion. The FA and MD were generated using DSI studio~\cite{dsistudio} from each of the reconstruction methods. 
\begin{table}[!ht]
    \footnotesize
    \begin{centering}
    \caption{\label{table 1} The Reconstruction Quality of the proposed method and the traditional CS method. Both NMSE and PSNR were calculated with fully sampled k-space  and q-space data (89 diffusion direction) as ground truth.}
    \resizebox{1.\columnwidth}{!}{\begin{tabular}{c|c|c|c|c|c}  \hline \hline
        AF    & & 2 & 4 & 6  & 8  \tabularnewline \hline \hline
        NMSE &  CS method & 26.89  & 45.33 & 62.38 & 99.06    \tabularnewline
        $\times$$10^{-13}$    & Proposed method  & \textbf{1.23}  & \textbf{5.80} & \textbf{6.57} & \textbf{9.13}\tabularnewline \hline
        PSNR &    CS method & 59.94  & \textbf{ 57.67} & 51.77 & 44.28    \tabularnewline 
          ($dB$)  & Proposed method  & \textbf{63.36}  &  56.60 & \textbf{56.06} & \textbf{54.63} \tabularnewline \hline\hline
\end{tabular}}
        \par\end{centering}
    \end{table}
From Tab.~\ref{table 1}, we can see with the increase of undersampling factor; for example, when AF = 8, both NMSE and PSNR get worse in the case of traditional CS. The proposed KLR method has almost $10$ fold improvement from traditional CS. For space limitation reasons, we do not include the examples of reconstructed diffusion weighted images. However, we present FA and MD maps obtained from the reconstructed images. Fig.~\ref{fig:fig2} - Fig.~\ref{fig:plot 1} are obtained using  undersampling in the k-space only and Fig.~\ref{fig:plot 3} -  Fig.~\ref{fig:fig6} - are obtained using undersampling in both k-space and q-space. All ground truths were computed using fully sampled k-space and fully sampled q-space data. From Fig.~\ref{fig:fig3} - Fig.~\ref{fig:plot 1}, we can see the proposed method outperforms traditional CS method in terms of reconstruction of FA and MD when we undersampled in k-space only. 

From  Fig.~\ref{fig:plot 3}-\ref{fig:fig6}, we see that when we have undersampling in both k-space and q-space performance of CS method drastically deteriorates while KLR still captures the correlation among the DWIs and have better estimation of FA and MD.  The error distribution plots in Fig.~\ref{fig:plot 1} and Fig.~\ref{fig:plot 3} shows the accuracy and distribution of errors in the estimation of FA and MD, obtained from the proposed KLR method and conventional CS method (Blue:KLR, Red:CS). The plots were generated using fully sampled k-space as reference. 
\section{Conclusion}
In this work, we have proposed an improved Kernel Low Rank method to reconstruct 3D dMRI images from undersampled data. The proposed method has shown superior performance compared to traditional CS methods. We have done extensive experiments and proved the robustness of KLR method in terms of NMSE, PSNR and several important map like FA and MD generation. Our future work will focus on exploring other non-linear low rank methods to improve the reconstruction.

\bibliographystyle{IEEEbib}
\bibliography{strings,refs}
\end{document}